\documentclass[twocolumn,prd,nofootinbib,aps,floats,floatfix,amsmath,amssymb,secnumarabic]{revtex4-1}

\usepackage{amssymb}

\usepackage{amsmath}
\usepackage{verbatim}
\usepackage{graphicx}
\usepackage[usenames]{color}
\usepackage{psfrag}
\usepackage{epstopdf}
\usepackage{multirow}
\usepackage[unicode]{hyperref}

\newcommand{\gd}{G_{\mbox{\tiny{RN}}}} 
\newcommand{\gr}{G_{\mbox{\tiny{R}}}}
\newcommand{\gn}{G_{\mbox{\tiny{N}}}}

\newcommand{\bb}{{\rm {{B}}}}
\newcommand{\mm}{{\rm {{M}}}}
\newcommand{\hh}{\mathcal{H}}
\newcommand{\nn}{\mathcal{N}}
\newcommand{\wmap}{\textit{WMAP}}
\newcommand{\Pl}{\textit{Planck}}

\newbox\tablebox    \newdimen\tablewidth
\def\leaderfil{\leaders\hbox to 5pt{\hss.\hss}\hfil}
\def\endtable{\tablewidth=\columnwidth 
    $$\hss\copy\tablebox\hss$$
    \vskip-\lastskip\vskip -2pt}

\def\tablenote#1 #2\par{\begingroup \parindent=0.8em
    \abovedisplayshortskip=0pt\belowdisplayshortskip=0pt
    \noindent
    $$\hss\vbox{\hsize\tablewidth \hangindent=\parindent \hangafter=1 \noindent
    \hbox to \parindent{\sup{\rm #1}\hss}\strut#2\strut\par}\hss$$
    \endgroup}
\def\doubleline{\vskip 3pt\hrule \vskip 1.5pt \hrule \vskip 5pt}

\begin{document}

\title{How does pressure gravitate?\\  
Cosmological constant problem confronts observational cosmology}

\author{Ali Narimani}
\email{anariman@phas.ubc.ca}
\affiliation{Department of Physics \& Astronomy,
University of British Columbia,
6224 Agricultural Road,
Vancouver, BC, V6T 1Z1  Canada} 
\author{Niayesh Afshordi}
\email{nafshordi@pitp.ca} 
\affiliation{Department of Physics \& Astronomy,
University of Waterloo, 
Waterloo, ON, N2L 3G1, Canada}
\affiliation{Perimeter Institute for Theoretical Physics, 
31 Caroline St. N., Waterloo, ON, N2L 2Y5, Canada}
\author{Douglas Scott}
\email{dscott@phas.ubc.ca}
\affiliation{Department of Physics \& Astronomy,
University of British Columbia,
6224 Agricultural Road,
Vancouver, BC, V6T 1Z1  Canada}

\begin{abstract}
An important and long-standing puzzle in the history of modern physics is the
gross inconsistency between theoretical expectations and cosmological
observations of the vacuum energy density, by at least 60 orders of magnitude,
otherwise known as the {\it cosmological constant problem}.  A characteristic
feature of vacuum energy is that it has a pressure with the same amplitude,
but opposite sign to its energy density, while all the precision tests of
General Relativity are either in vacuum, or for media with negligible pressure.
Therefore, one may wonder whether an anomalous coupling to pressure might be
responsible for decoupling vacuum from gravity.  We test this possibility in
the context of the {\it Gravitational Aether\/} proposal, using current
cosmological observations, which probe the gravity of relativistic pressure
in the radiation era.  Interestingly, we find that the best fit for anomalous
pressure coupling is about half-way between General Relativity (GR), and
Gravitational Aether (GA), if we include {\it Planck\/} together with
{\it WMAP\/} and BICEP2 polarization cosmic microwave background (CMB)
observations.  Taken at face value, this data combination excludes both GR
and GA at around the $3\,\sigma$ level.  However, including higher resolution
CMB observations (``highL'') or baryonic acoustic oscillations (BAO) pushes
the best fit closer to GR, excluding the Gravitational Aether solution to the
cosmological constant problem at the 4--$5\,\sigma$ level.  This constraint
effectively places a limit on the anomalous coupling to pressure in the
parametrized post-Newtonian (PPN) expansion, $\zeta_4 = 0.105 \pm 0.049$
(+highL CMB), or $\zeta_4 = 0.066 \pm 0.039$ (+BAO).  These represent the
most precise measurement of this parameter to date, indicating a mild
tension with GR (for $\Lambda$CDM including tensors, with $\zeta_4=0$),
and also among different data sets.  

\end{abstract}

\maketitle

\section{Introduction}

One of the most immediate puzzles of quantum gravity (i.e., applying the rules
of quantum mechanics to gravitational physics) is an expectation value for the
vacuum energy that is 60--120 orders of magnitude larger than its measured
value from cosmological (gravitational) observations. This is known as the
(now old) {\it cosmological constant problem\/} \cite{1989RvMP...61....1W}, and
has been thwarting our understanding of modern physics for almost a century
\cite{2002gr.qc.....8027S}.   The discovery of late-time cosmic acceleration
\cite{1998AJ....116.1009R, 1999ApJ...517..565P}, added an extra layer of
complexity to the puzzle, showing that the (gravitational) vacuum energy,
albeit tiny, is non-vanishing (now dubbed, the {\it new\/} cosmological constant
problem).

Gravitational Aether (GA) theory is an attempt to find a solution to the
{\it old\/} cosmological constant problem \cite{2008arXiv0807.2639A, aslan},
i.e., the question of
why, in lieu of fantastic cancellations, the vacuum quantum fluctuations do
not appear to source gravity.
The approach is to stop the quantum vacuum from gravitating by modifying our
theory of gravity, as we describe below.
In this way the (mean density of ) quantum fluctuations will have no dynamical
effect in astrophysics or cosmology
(see \cite{zeldovich} for one of the very first steps and
\cite{zhitnitsky} for an alternative but related attempt for solving the problem).

Although GA is a very specific proposal for modifying gravity, it may serve
as an example of more general theories.  As we will see below, a generalized
version may represent a broader class of theories in which the gravitational
effects of pressure (and including anisotropic stress) might be different
from those of GR.

It is important to be clear that this theory does not have any solution for
the ``new'' cosmological constant problem, i.e., the empirical existence of
a small vacuum energy density which now dominates the energy budget of the
Universe, driving the accelerated expansion and making the geometry of space
close to flat.  Hence in GA theory it is assumed that the vacuum quantum
fluctuations
(the old problem) and the small but non-zero value of $\Lambda$
(the new problem) are two separate phenomena that should be explained 
independently (but see \cite{2009PhRvD..80d3513P}). 

The Einstein field equations in the GA theory (in units with $c=1$, 
and with metric signature=(--+++))
are modified to
\begin{eqnarray}
(8\,\pi\,\gr)^{-1}(G_{\mu \nu}+\Lambda g_{\mu \nu})&=& 
T_{\mu \nu} -\dfrac{1}{4}T^\alpha_\alpha g_{\mu \nu} + T'_{\mu \nu},
 \label{eqn1} \\
{\rm with}\ T'_{\mu \nu}&=& p'(u'_\mu u'_\nu + g_{\mu \nu}). \label{tprime}
\end{eqnarray}
Most significantly, the second term on the right hand side of Eq.~\ref{eqn1}, 
$-\frac{1}{4}T^\alpha_\alpha g_{\mu \nu}$, solves the old cosmological constant problem
by cancelling the effect of vacuum fluctuations in the energy momentum tensor. 
The third term, $T'_{\mu \nu},$ is then needed to make the field
equations consistent, and is dubbed {\it gravitational aether}.  

The form used for $T'_{\mu \nu}$ in Eq.~\ref{tprime} is a convenient choice,
but is probably not unique, although it is limited by phenomenological and
stability constraints \cite{2008arXiv0807.2639A}.
However, $p'$ and $u'_\mu$, the pressure and four velocity unit vector of the aether, 
are constrained through the terms in the energy-momentum tensor by applying
the Bianchi 
identity and the assumption of energy-momentum conservation, i.e.,
\begin{equation}
\nabla^\mu T'_{\mu \nu} = \dfrac{1}{4} \nabla_\nu T. \label{constraint}
\end{equation}

The only free constant of this theory, as in General Relativity (GR), is $\gr$, although, as we will see, 
this is not the same as the usual Newtonian gravitational constant, $\gn$. In addition,
of course, there are parameters describing the constituents in the various tensors, i.e., the 
cosmological parameters. In cosmology, the 
energy-momentum tensor, $T_{\mu \nu}$, consists of the conventional fluids, i.e., radiation, 
baryons, and cold dark matter, plus a contribution due to vacuum fluctuations,
\begin{eqnarray}
T_{\mu \nu}&=& T^{{\rm R}}_{\mu \nu}+T^{{\rm B}}_{\mu \nu}+T^{{\rm C}}_{\mu \nu}
+\rho_{\rm vac} g_{\mu \nu} , \\
T& \equiv& T^\alpha_\alpha \, = \, - (\rho^{\rm B}+\rho^{\rm C}) + 4\,\rho_{\rm vac},
\end{eqnarray}
where neutrinos are included as part of radiation, and their mass is set to zero in this paper. 

Equations~\ref{eqn1} and \ref{tprime} that describe GA are drastic
modifications of GR with no additional tunable parameter. Therefore, one may
wonder whether GA can survive all the precision tests of gravity that have
already been carried out. These tests
are often expressed in terms of the parameterized post-Newtonian (PPN)
modifications of GR, which are expressed in terms of 10 dimensionless PPN
parameters \cite{2014arXiv1403.7377W}.  While these parameters do not capture
all possible modifications of GR, they are usually sufficient to capture
leading corrections to GR predictions in the post-Newtonian regime
(i.e., nearly flat space-time with non-relativistic motions), in lieu of new
scales in the gravitational theory.  It turns out that only one PPN parameter,
$\zeta_4$, which quantifies the anomalous coupling of gravity to pressure has
not been significantly constrained empirically, as the existing precision
tests only probe gravity in vacuum, or for objects with negligible pressure.
Indeed, since only the sourcing of gravity is modified in GA, the vacuum
gravity content is identical to GR, and the only PPN parameter that deviates
from GR is $\zeta_4=1/3$ (as opposed to $\zeta_4=0$ in GR)
\cite{2008arXiv0807.2639A, aslan}.   
 
The idea that $\zeta_4$ could be non-zero runs contrary to the conventional
wisdom that relates gravitational
coupling to pressure on the one hand, to the couplings to internal and kinetic
energies on the other \cite{1976ApJ...204..224W}, both of which are already
significantly constrained by
experiments.  However, this expectation is based on the assumption that the
average gravity of a gas of interacting point particles, is the same as the
gravity of a perfect fluid that is obtained by coarse-graining the particle
gas \footnote{This would not be the case in the GA theory, since the aether
tracks the motion of individual particles, due to the constraint
of Eq.~\ref{constraint}.  Therefore, the nonlinear back-reaction of the
motion of the aether would be lost in the coarse-grained perfect fluid.}.
This connects with the whole issue of the assumption of the continuum
approximation for cosmological fluids, where the particle density is low, so
that the average distance between particles is a macroscopic scale.
Gravity is only well-tested on scales $\gtrsim 0.1\,$mm
\cite{2007PhRvL..98b1101K}, which are larger than the distance between
particles in most terrestrial or astrophysical precision tests of gravity.
Therefore, there is no guarantee that the same laws of gravity apply to
microscopic constituents of the continuous media in which gravity is currently
tested.  Indeed, GA could only be an effective theory of gravity above some
scale $\lambda_{\rm c} \lesssim 0.1$ mm, implying that sources of
energy-momentum on the right hand sides of Eqs.~\ref{eqn1} or \ref{constraint}
should be coarse-grained on scale $\lambda_{\rm c}$.


At first sight, it might appear that the dependence of gravitational
coupling on pressure signals a violation of weak and/or strong
equivalence principles (WEP and/or SEP). However, WEP is explicitly imposed
in GA, as all matter components couple to the same metric. Moreover, SEP is
so far only tested for gravity in vacuum (e.g.\ point masses in the solar
system), where GA is equivalent to GR, as aether is not sourced, and thus
vanishes (in lieu of non-trivial boundary conditions; see e.g. \cite{2009PhRvD..80d3513P}).

What goes against one's intuition in the case of the GA modification of
Einstein
gravity, compared to e.g.\ scalar-tensor theories, is that even in the
Newtonian limit, comparable effects come from the change in couplings {\it
and\/} the gravity of the energy/momentum of the aether. In contrast, the
additional fields in the usual modified gravity theories carry little
energy/momentum in the Newtonian regime, while they could modify couplings
by order unity. If the change in the gravitational mass (due to the
dependence of $G$ on the equation of state) is by the same factor as the
change in energy/momentum (due to the additional terms on the RHS of
Einstein equations), then the ratio of gravitational to inertial mass
remains unchanged.

A more intuitive picture might be to consider aether (minus the trace term)
as an exotic fluid {\it bound\/} to matter, similar to an electron gas for
example, within ordinary GR.  Like the electron gas, the effect will be to modify the gravitational field source, by the amount of energy/momentum in the exotic fluid. However, unlike the electron gas, the non-gravitational energy/momentum exchange between matter and the exotic fluid is tuned to zero, which ensures WEP, at least at the classical level.    Moreover, the action-reaction principle (Newton's 3rd law) for gravitational forces should include the momentum in, and interaction with the exotic fluid.  


Another conceptual issue with Eqs~\ref{eqn1}--\ref{tprime} is that, at least
to our knowledge, they do not follow from an action principle.  However, an
action principle may not be necessary (or even possible) for a low energy
effective theory, such as in the case of Navier-Stokes fluid equations, even
if the fundamental theory does follow from an action principle.  Given the
severity of the cosmological constant problem, it seems reasonable that we
might be prepared to relax requirements that are not absolutely necessary
for a sensible effective description of nature. 

There are two obvious places in the Universe to look for the gravitational
effect of relativistic pressure, and thus constrain $\zeta_4$:
\begin{enumerate}
 \item The first situation involves compact objects, particularly the internal
 structure of neutron stars \cite{Schwab2008,2011PhRvD..84f3011K}.
 While, in principle, mass and radius measurements of neutron stars can be used  
 to constrain $\zeta_4$, at the moment the constraints are almost completely
 degenerate with the uncertainty in the nuclear equation of state (not to
 mention other observational systematics).  However, future observations of
 gravitational wave emission from neutron star mergers (e.g., with Advanced
 LIGO interferometers) might be able to break this degeneracy
 \cite{2011PhRvD..84f3011K}.  It may also be possible to develop tests that
 probe near the hot accretion disks of black holes or during the formation
 of compact objects in supernova explosions.
 
 \item The second situation is the matter-radiation transition in the early
 Universe.  Ref.~\cite{Rappaport2008} studied constraints arising from the
 big bang nucleosynthesis epoch.  However, more precise measurements come from
 various cosmic microwave background
 (CMB) anisotropy experiments, such as the {\it Wilkinson Microwave Anisotropy
 Probe\/} ({\it WMAP\/}) \cite{wmap}, {\it Planck\/} \cite{PlanckMission}, the Atacama
 Cosmology Telescope (ACT) \cite{act}, and the South Pole Telescope (SPT)
 \cite{spt}, amongst other cosmological observations.  The constraints on GA
 were studied in detail in Ref.~\cite{aslan}, with the data sets available at
 that time.  While GA might arguably ease tension among certain observations,
 such as the Ly-$\alpha$ forest, primordial Lithium abundance, or earlier
 ACT data,
 it was discrepant with others, such as Deuterium abundance, SPT data,
 or low-redshift measurements of cosmic geometry.  The aim of this paper
 is to carefully revisit these tensions in observational cosmology, in
 light of the significant advances within the past three years.   
 \end{enumerate}

With this introduction, in Sec.~\ref{sec:eqs}, we move on to derive the equations for the cosmological background, as well as linear perturbations, in the GA theory. Similar to Ref.~\cite{aslan}, we use the Generalized Gravitational Aether (GGA) framework, which interpolates between GR and GA, to quantify the observational constraints. This framework depends on the ratio of the gravitational constant in the radiation and matter eras, $\gr/\gn = 1+\zeta_4$, which is $1+\frac{1}{3} =\frac{4}{3}$ ($1+0 =1$) for GA (GR).  Sec.~\ref{sec:constraints} discusses our numerical implementation of the GGA equations, and the resulting constraints from different combinations of cosmological data sets, some of which appear to exclude GA at the 4--5$\,\sigma$ level, while others are equally (in)consistent with GR or GA at about the $3\,\sigma$ level. Finally, Sec.~\ref{sec:sum} summarizes our results, discusses various open questions, and highlights avenues for future inquiry. 

\section{Equations of motion at the background and perturbative level} \label{sec:eqs}

Baryons, radiation, and cold dark matter can be considered as perfect fluids with simple
equations of state, $p = w \rho$, at the background level. The following $p'$ and $u'$ 
will solve Eqs.~\ref{eqn1}--\ref{constraint} in this case:
\begin{eqnarray}
p'&=& \sum\limits_i \dfrac{(1+w^i)(3\,w^i-1)}{4} \rho^{i} ; \label{pprime} \\
u'_\mu&=& \sum\limits_i \dfrac{ (1+w^i)(1-3\,w^i)}{2} u^{i}_{\mu}
. \label{uprime}
\end{eqnarray}
Here, ``$i$'' stands for either baryons, radiation, cold dark matter, or vacuum fluctuations.
Based on Eqs.~\ref{pprime} and \ref{uprime}, $p'\, =\, -(\rho^{\rm B}+
\rho^{\rm C})/4$, and $u'_{\mu}\, =\, u^{\rm C}_{\mu}$ at the background level.
Substituting these relations back into Eq.~\ref{eqn1}, the field equations
will take the following form in terms of the conventional fluids in $T_{\mu \nu}$:
\begin{equation}
(8\,\pi)^{-1}(G_{\mu \nu}+\Lambda g_{\mu \nu}) = 
\gr \,T^{{\rm R}}_{\mu \nu} + \dfrac{3}{4} \, \gr \,
 (T^{{\rm B}}_{\mu \nu}+T^{{\rm C}}_{\mu \nu}). \label{back}
\end{equation}

One of the clearest testable predictions of this theory is that space-time reacts differently
to matter and to radiation: a spherical ball full of relativistic matter curves the space-time more
than a spherical ball of non-relativistic substance (of the same size and density). Defining $\gr \equiv 4\, \gn/3$, where 
$\gn$ is the usual Newtonian gravitational constant, and using the FRW metric,
$ds^2 = a^2(- d \tau^2 + d \textbf{x}^2)$,
the Friedmann equation in the GA theory will be:
\begin{equation}
\mathcal{H}^2 = \dfrac{8 \pi \gn a^2}{3} (\rho + \dfrac{1}{3}\rho^{\rm R}) \quad , \quad
\rho = \rho^{\rm R} + \rho^{\rm B} + \rho^{\rm C} + \rho^{\rm \Lambda}. \label{hub1}
\end{equation}
$\hh$ is defined as $\dot{a}/a$ here, and a dot represents a derivative with respect to the conformal time, $\tau$.

The Friedmann equation can be used to calculate the predictions of the theory for big bang 
nucleosynthesis (BBN) (see e.g., Ref.~\cite{aslan}). Although the different effective value of $G$ in the
early Universe means that the BBN predictions are different from the standard model,
uncertainties in the consistency of the light element abundances suggest that the comparison with data
cannot be considered as fatal for the theory. Therefore, one needs to go one step further and calculate 
the first-order perturbations to determine the predictions for observables such as the
CMB anisotropies, or the matter power spectrum. 

Before dealing with the perturbations, it is worth noticing that the GA theory 
can be treated as a special case of a more general framework. We shall call this the Generalized 
Gravitational Aether (GGA), which has the following field equations:
\begin{equation}
(8\,\pi)^{-1}(G_{\mu \nu}+\Lambda g_{\mu \nu}) = 
\gr T_{\mu \nu} -\gd T^\alpha_\alpha g_{\mu \nu} + 4\, \gd T'_{\mu \nu} . \label{GGA}
\end{equation}
Here $\gd \equiv \gr - \gn = \zeta_4 \gn$ is the difference in gravitational constants between radiation 
and matter.
$\gr$ and $\gn$ are both free constants and one will recover the gravitational aether by 
setting $\gr = 4 \gn/3$. General relativity is also a special case of GGA, with $\gd = 0$. 
The Friedmann equation in GGA will be: 
\begin{equation}
\mathcal{H}^2 = \dfrac{8 \pi a^2}{3} (\gn \, \rho + \gd \, \rho^{\rm R}) .
\end{equation}
Using GGA as a framework, we then have a family of models, parameterized by
$\zeta_4$, with $\zeta_4=0$ corresponding to GR and $\zeta_4=1/3$ being GA.

It is fairly straightforward to calculate the perturbation equations in the general (GGA) 
framework, which will then contain GA and GR as special cases.
We will use the cold dark matter gauge (see e.g., Ref.~\cite{ma}) with the following metric for the first 
order perturbations:
\begin{eqnarray}
& & ds^2 =  a^2(\tau) \left[ -d\tau^2 + ( \delta_{ij} + h_{ij} ) dx^idx^j \right] ; \nonumber   \\
& & h_{ij} =  \int d^3k e^{i\vec{k}.\vec{x}} \left[ \hat{k}_i \hat{k}_j h(\vec{k},\tau)
+\left(\hat{k}_i \hat{k}_j-\dfrac{1}{3} \delta_{ij} \right) 6 \eta(\vec{k},\tau) \right] ; \nonumber \\
&& \vec{k}  = k\hat{k} \label{metric} . 
\end{eqnarray}
We will also use the following definitions for the perturbation parts of the energy momentum tensor:
\begin{eqnarray}
\delta T^0_0&=& -\delta \rho ; \\
\delta T^0_i&=& (\bar{\rho}+\bar{p}) \,V_i ; \\
\delta T^i_i&=& 3\,\delta p ; \\
\mathcal{D}_{ij} \delta T^{ij}&=&  (\bar{\rho}+\bar{p})\Sigma .
\end{eqnarray}
The barred variables refer to background quantities and $\mathcal{D}_{ij}$ is  
defined as $\hat{k}_i \hat{k}_j-\dfrac{1}{3} \delta_{ij}$. Once again, the fluids in 
$T_{\mu \nu}$ are baryons, cold dark matter, radiation, and vacuum quantum fluctuations.
We will follow the conventions of Ref.~\cite{aslan} and define the  
perturbations in the aether density and four velocity as
\begin{equation}
\delta p' = p' - \left(-\dfrac{\rho^{\rm M}}{4}\right) \ , \quad
\delta u'_\mu = u'_\mu - u^{\rm C}_\mu.
\end{equation}
Here $\rho^{\rm M}$ is the total matter density, i.e., baryons plus cold dark matter, 
and the quantities $\rho^{\rm M}$ and $u^{\rm C}_\mu$ consist of both their background 
and perturbation parts. Using the above definitions and the metric defined in
Eq.~\ref{metric}, we obtain four equations of motion from the GGA field equations:
\begin{eqnarray}
& & k^2  \eta - \frac{1}{2} \, \hh \, \dot{h} =
- 4 \, \pi \, \gn \, a^2 \delta \rho + k^2 \, A(k,\tau) ; \label{first}  \\
& & k  \dot{\eta} = 4 \, \pi \, \gn \, a^2 (\bar{\rho}+\bar{p}) \,  V 
+ k^2  B(k,\tau) ; \label{second} \\
& & \ddot{h} + 2\,\hh \, \dot{h} - 
2\, k^2 \, \eta = - 24\, \pi \, \gn \, a^2 (\delta P) 
+ k^2\, C(k,\tau) ; \label{third} \\ 
& & \ddot{h} + 6 \ddot{\eta}  + 2\, \hh \,
(\dot{h} + 6 \dot{\eta}) - 2\, k^2 \eta =
-24 \pi  \gn \, a^2 (\bar{\rho}+\bar{p})\Sigma \nonumber \\
& &  \qquad \qquad \qquad \qquad \qquad + k^2 D(k,\tau).  \label{last}
\end{eqnarray}
The four functions, $\{A,B,C,D\}$, are
\begin{eqnarray}
& & A(k,\tau) = \dfrac{-4\, \pi \, \gd \, a^2 \, \delta \rho^{\rm R}}{k^2}, \label{Aa} \\
& & B(k,\tau) = \dfrac{4\,\pi \, \gd \, a^2 \,(i\,k^i \delta T^0_i - 
               \bar{\rho}^{\rm M}\, \omega)}{k^3},\label{Bb} \\
& & C(k,\tau) = \dfrac{-8\, \pi \, \gd \, a^2 (\delta \rho^{\rm R}+12\,\delta p')}{k^2},  \\
& & D(k,\tau) = \dfrac{-24 \, \pi \, \gd \, a^2 \mathcal{D}_{ij} \delta T^{\rm R}_{ij} }{k^2} . 
\label{Dd}
\end{eqnarray}
Here $\omega$ is defined as the divergence of the aether four velocity perturbation: 
$\omega \equiv i k^i \delta u'_i/a$. One can equally use Eqs.~\ref{first} to \ref{last}, or use 
Eq.~\ref{constraint} to derive the following two constraints for the aether parameters:
\begin{eqnarray}
3\, \dfrac{\hh}{a}\,\partial_\tau(a\,\omega) + k^2 \, \omega&=& k^2 \dfrac{\bar{\rho}^{\bb}_{0}}
                                                    {\bar{\rho}^\mm_{0}}\, \theta^\bb ;
                                                    \label{omega} \\
\delta p'&=& \dfrac{\bar{\rho}^\mm}{12\, \hh} ( \omega - \dfrac{\bar{\rho}^\bb_{0}}
                                                    {\bar{\rho}^\mm_{0}}\, \theta^\bb
                                                    ) .
\end{eqnarray}
Here $\bar{\rho}^{{\rm B}}_{0}$ and $\bar{\rho}^{{\rm M}}_{0}$ are the current background density 
in baryons and matter, respectively, and $\theta^{\rm B}$ is the divergence of the
baryon velocity perturbation: $\theta \equiv i k^i\, V^{{\rm B}}_{i}$. At very early times, when
$k \ll \hh$, one can ignore the right hand side of Eq.~\ref{omega}, together with the $k^2\,\omega$
factor on the left hand side. The initial condition for the divergence should therefore be deduced 
from 
\begin{equation}
\dot{\omega} + \hh\, \omega = 0 \, . \label{Om_ini}
\end{equation}
Any non-zero initial condition on $\omega$ will be damped as $a^{-1}$, and it is therefore
reasonable to assume the  initial condition $\omega =0 $ at all scales. It is also interesting to 
notice that, since we are using the cold dark matter gauge, $\omega$ will once again be washed 
out for very large scales, $k \ll \hh$,
at late times when baryons fall into the potential well of the cold dark matter particles 
and start co-moving with them. 

The physical meaning of the four modifying terms, $\{A,B,C,D\}$, is explained in Ref.~\cite{mog} for 
an even more general theory. In short, the second term and the time derivative of the first term will 
act as driving forces for matter overdensities, while the second term and the time derivative of the 
fourth term are important in the integrated Sachs-Wolf (ISW)~\cite{isw} effect.

We will confront the GGA theory with cosmological observations in the next section.

\section{Cosmological constraints on GGA} \label{sec:constraints}
We have modified the cosmological codes {\tt CAMB} \cite{camb} and
{\tt CosmoMC} \cite{cosmomc} 
in order to test the predictions of GGA
against cosmological data. Before confronting the theory with data, it is necessary to make
sure that the codes are internally consistent and error-free. We will list a number of consistency 
checks we have made on {\tt CAMB} in the next subsection, and then report the constraints on the GGA 
parameter.

\subsection{Consistency checks on CAMB} \label{Sec:CAMB_Cons}
One of the relatively trivial tests on the modified {\tt CAMB} code is that it should reproduce 
the $C_\ell$s of the non-modified code after setting $\gd = 0$. The next obvious thing is a test at 
the background level.
The GA theory is completely degenerate at the background level with a GR model that has one 
third more radiation (see Eq.~\ref{hub1}). In the standard picture each light neutrino species
adds $0.23$ times as much radiation as the photons. Therefore, the following models should result
in exactly the same $a(\tau)$ and $\hh(\tau)$ functions: 
${\cal B}1:=\{\gd = 1/3 \gn, \nn_{\rm eff}=3.04\}$\footnote{Here ${\cal B}$ is
for background, and ${\cal P}$ will be for perturbations.}
and ${\cal B}2:=\{ \mbox{GR with}\, \, \nn_{\rm eff}=5.54 \}$, where $\nn_{\rm eff}$ is the effective number 
of light neutrinos.

\begin{figure}[h]
\includegraphics[scale = 0.45]{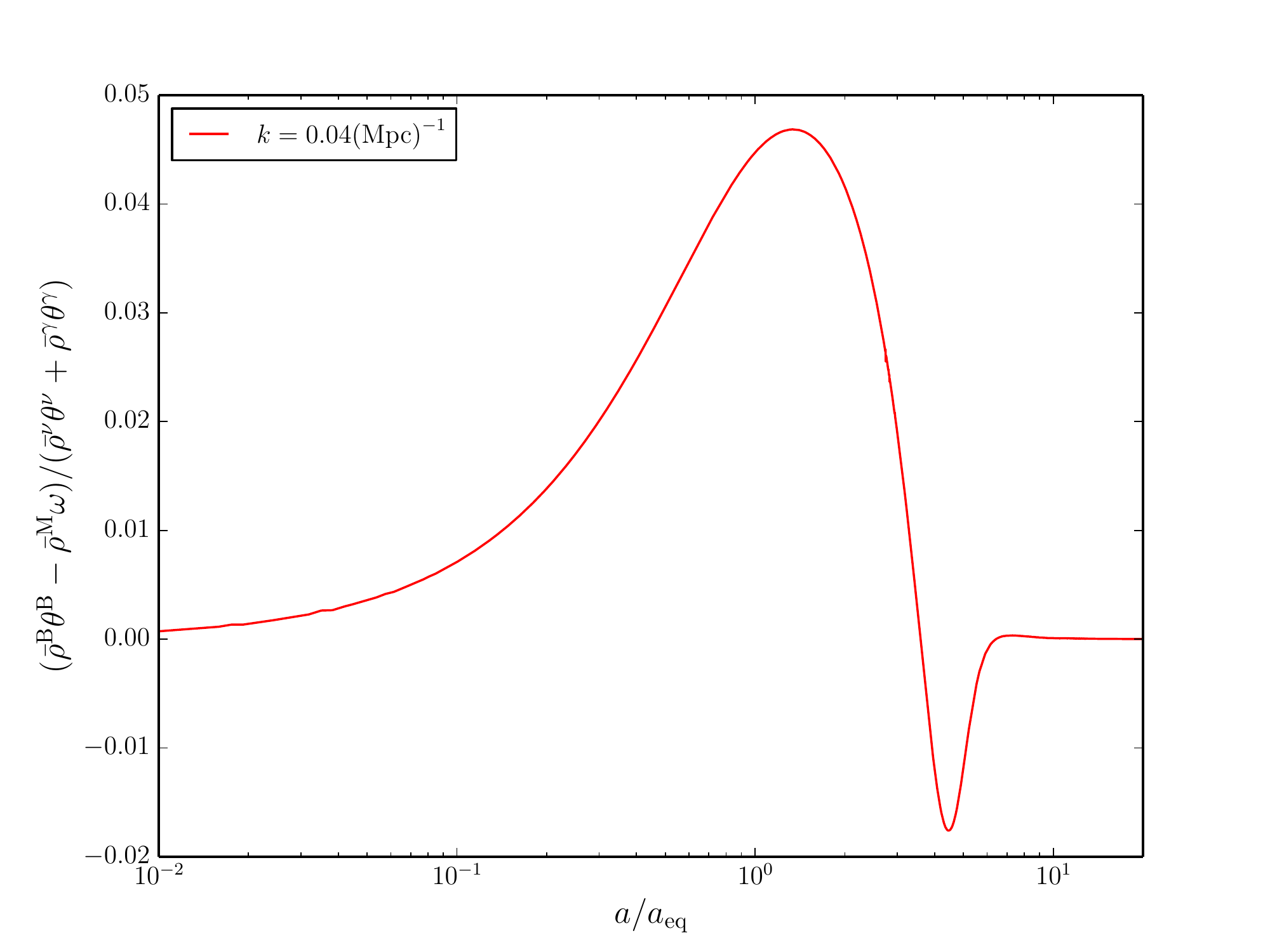}
\caption{\label{fig:rad_nonrad} Effects of non-radiation modifying terms  
compared to the effects of radiation, at $k= 0.04$ Mpc$^{-1}$ (see Eq.~\ref{bb2}). This shows that the non-radiation 
terms are smaller by more than a factor of $20$.} 
\end{figure}

The effect of GGA at the perturbation level is evident through the four modifying functions
$\{A,B,C,D\}$. Using constraint equations such as Eq.~\ref{constraint}, one can see that 
the functions $\{C,D\}$ are linear combinations of the first two functions $\{A,B\}$ and 
their time derivatives. Therefore, the two functions $\{A,B\}$ are sufficient for tracing the 
perturbative effects of GGA. Between the two, $A$ is purely dependent on radiation at the 
perturbation level (see Eq.~\ref{Aa}). Looking closer at $B$ in Eq.~\ref{Bb} we see that

\begin{equation}
k^3 B = \dfrac{4}{3}\,\pi\, \gd\, a^{2} \left( 3 \Delta \omega +4\,{\it \bar{\rho}}^{{\nu
}}\theta^{{\nu}}+4\,{\it \bar{\rho}}^{{\gamma}}\theta^{{\gamma}} \right) . \label{bb2}
\end{equation}
Here $\Delta \omega$ is defined as $(\,{\it \bar{\rho}}^{{\rm B}} \, 
\theta^{{\rm B}}-\,{\it \bar{\rho}}^{{\rm M}} \, \omega)$, which is proportional to the time 
derivative of $\omega$, according to Eq.~\ref{omega}, and is therefore smaller than the radiation 
terms (see Fig.~\ref{fig:rad_nonrad}). $\theta^{\nu}$ and $\theta^\gamma$ are the neutrino and 
photon first moments, respectively \cite{ma}.

Putting this together, we find that the GGA effects are almost degenerate with extra radiation,
even when we consider perturbations.
However, the GGA-$\nn_{\rm eff}$ degeneracy does not hold exactly at the perturbation
level, since $\delta \rho^{\gamma}/\delta \rho^{\nu}$ and 
$\theta^{\gamma}/\theta^{\nu}$ are both time- and scale-dependent, contrary to the
previous case at the background level, where $\bar{\rho}^\gamma / \bar{\rho}^\nu$ was a 
constant number through time and for every scale. 

\begin{figure}[ht]
\includegraphics[scale = 0.45]{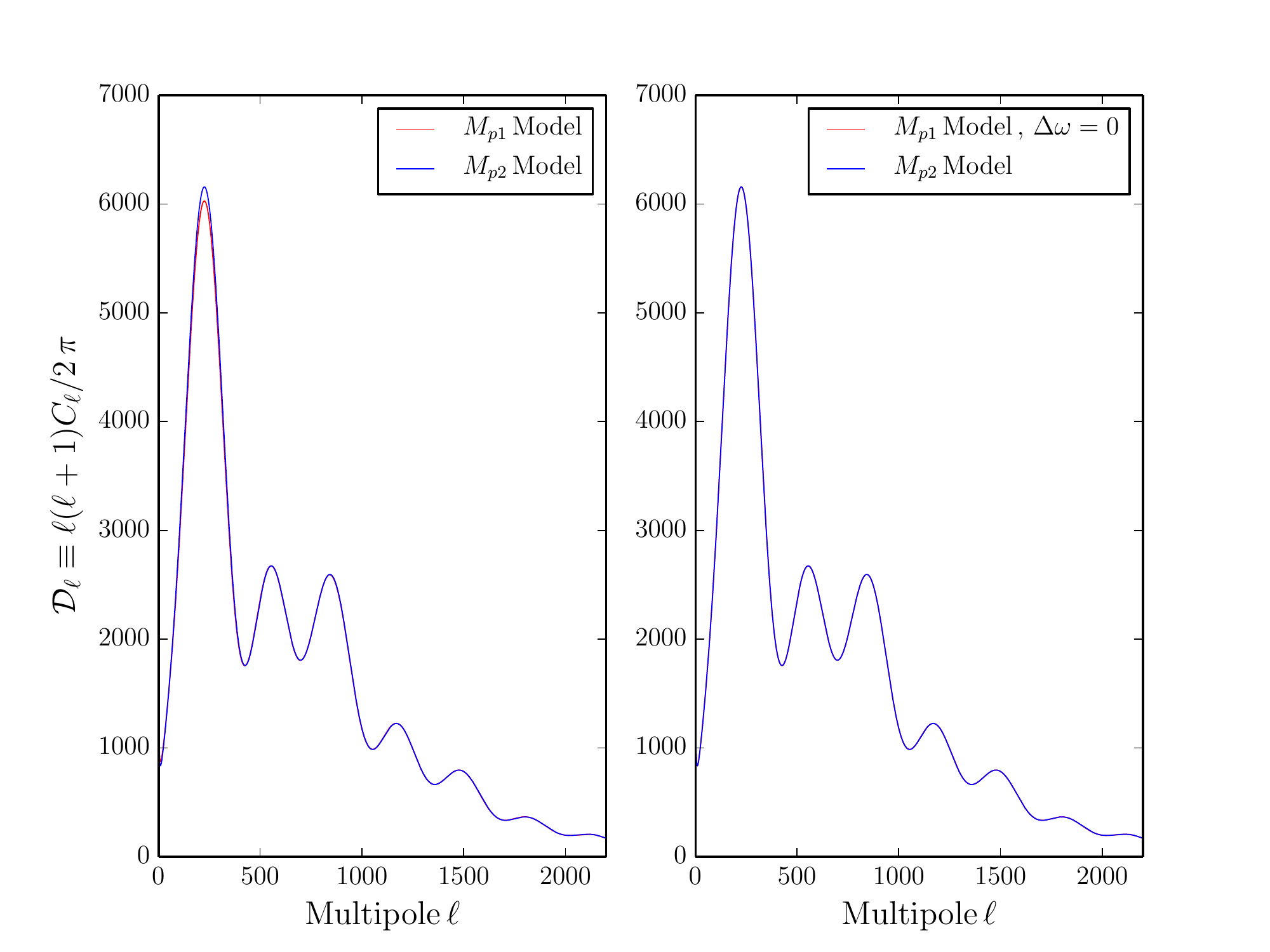}
\caption{\label{fig:final} Checking the code at the perturbation level by
comparing the matter power spectra for the models ${\cal P}1:=\{\gd = 1/3 \gn, \nn_{\rm eff}=3.04, T_0^4 = (2.7255)^4 \}$ 
and ${\cal P}2:=\{ \mbox{GR with}\, \nn_{\rm eff}= 4/3 \times 3.04,\, T_0^4 = 4/3 \times 
(2.7255)^4  \}$ (left panel). There is a small difference between the two
at around the first peak. This can be explained by considering the effects of a non-zero
$\omega$ (divergence of the aether four-velocity). The two models completely coincide with each other by setting
$\Delta \omega $ to zero in the ${\cal P}1$ model (right panel). }
\end{figure}

\begin{figure}[ht]
\includegraphics[scale = 0.45]{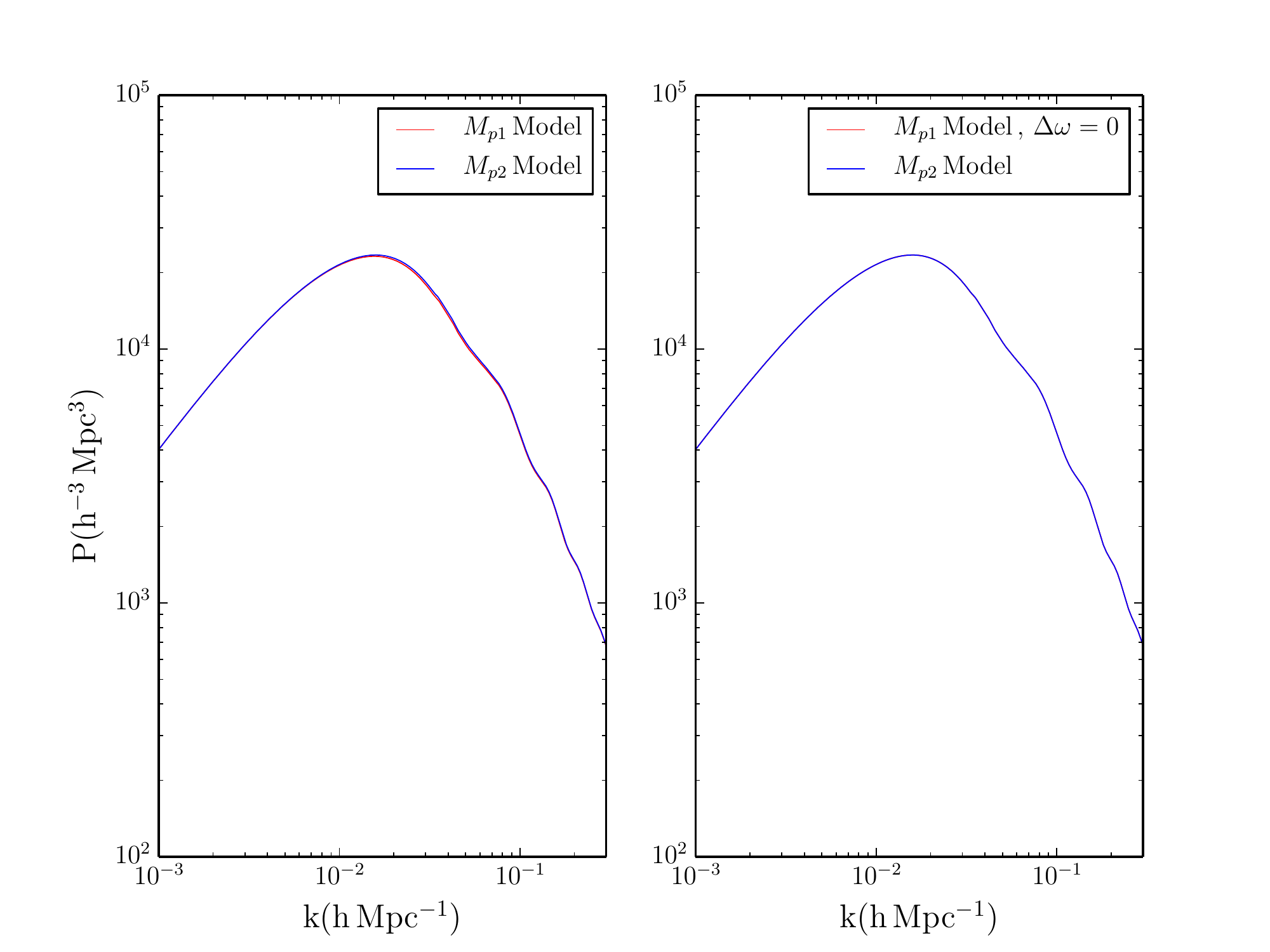}
\caption{\label{fig:final2} Checking the code at the perturbation level by
comparing the the models ${\cal P}1:=\{\gd = 1/3 \gn, \nn_{\rm eff}=3.04, T_0^4 = (2.7255)^4 \}$ 
and ${\cal P}2:=\{ \mbox{GR with}\ \nn_{\rm eff}= 4/3 \times 3.04,\, T_0^4 = 4/3 \times 
(2.7255)^4  \}$ (left panel). The small difference between the two
can be explained by considering the effects of a non-zero
$\omega$. The two models completely coincide with each other by setting
$\Delta \omega $ to zero in the ${\cal P}1$ model (right panel). }
\end{figure}

The final test of the modified {\tt CAMB} code is at the perturbation level when the
GA parameter, $\gd$, is not set to zero and GA is fully effective.
Based on the discussion in the preceding paragraphs the code
should produce the same $C_\ell$s for the 
following two models: ${\cal P}1:=\{\gd = 1/3 \gn, \nn_{\rm eff}=3.04, T_0^4 = (2.7255)^4\}$ and 
${\cal P}2:=\{ \mbox{GR with}\, \nn_{\rm eff}= 4/3 \times 3.04,\, T_0^4 = 4/3 \times (2.7255)^4 \}$,
when the non-radiation terms, i.e., terms in $\Delta \omega$ are ignored. 
We use the value of the present CMB temperature, $T_0$, from Ref.~\cite{fixsen}.
However, one needs to be careful while performing
this test, since $T_0$ appears in many parts of the code that are totally irrelevant to gravity 
(see Ref.~\cite{narimani} for a related discussion), e.g., the sound speed of the plasma before 
last scattering depends on the photon-to-baryon density 
ratio and hence on $T_0$.  Figs.~\ref{fig:final} and \ref{fig:final2} show
these tests of our calculations for GGA.
The right panel of these figures tests the code at the perturbation level, while the left panels
show the effect of non-radiation fluids on the CMB anisotropies and matter power spectra. Ignoring
the effect of non-radiation fluids, is crucial in reducing the GA Eqs.~\ref{eqn1}--\ref{constraint},
to Eq.\ref{back} at the perturbation level, and hence the $\gd$--$\zeta_4$ correspondence. 

\begin{figure}[ht]
\includegraphics[scale = 0.45]{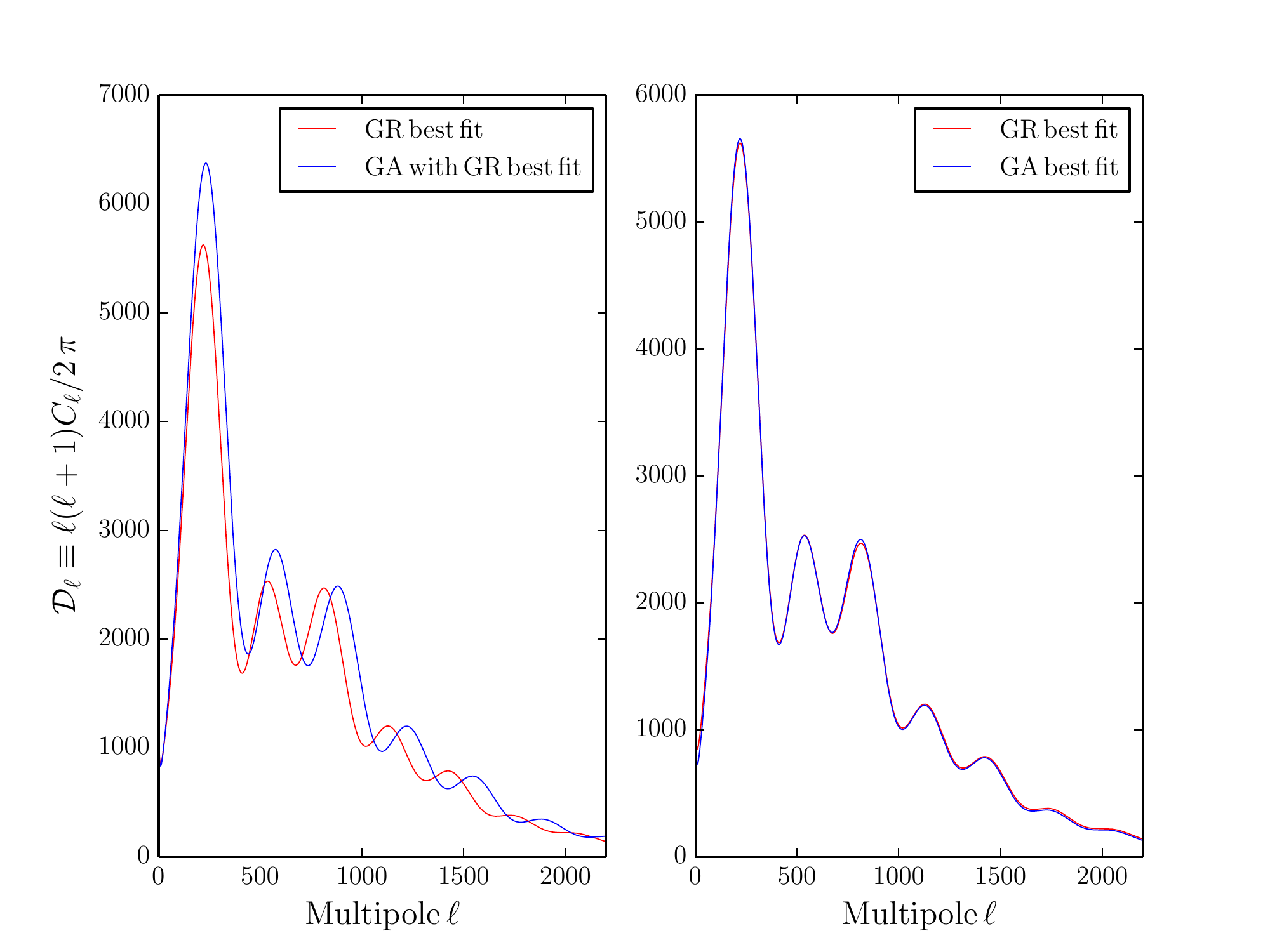}
\caption{\label{fig:cell} Comparing general relativity versus gravitational aether
predictions for the CMB power spectrum. The values of the input parameters for the left
panel are taken from the {\it Planck\/} analysis \cite{Planck}. The right panel compares the best-fit predictions of the
two theories with all cosmological parameters also allowed to vary. 
GA predicts less power at higher $\ell$s, as one can see from the right panel.}
\end{figure}

Figure~\ref{fig:cell} shows the CMB anisotropy power spectrum predictions from GR and GA
(with other GGA models interpolating between the two).
The input parameters of the left panel are the same for both theories and are taken from
Ref.~\cite{Planck}. We see on the left panel that the positions of the peaks 
are consistently shifted towards smaller scales,
i.e., higher $\ell$s. This is because the Universe is younger at recombination in the GA theory,
which in turn is due to having effectively more radiation at the background level of the GA theory
compared to GR. There is also an enhanced early ISW effect \cite{isw1} in the GA theory due 
the presence of the the two modifying functions, $B$ and $D$, as was explained before. 
This can be understood more intuitively using the fact that GA is effectively degenerate at 
the background and perturbation level with a GR model with one third more radiation. 
Since the ISW effect is proportional to $e^{-\tau}$ (where $\tau$ is the optical depth), 
and the time derivative of the metric potentials (that are non-zero only during the
matter radiation transition and at very late times),
then having more radiation in the Universe will delay the radiation to matter transition 
to later times with smaller $\tau$ and enhance the ISW effect.

In order for the GA model to match GR and hence fit the data, since there is a very good match 
between data and GR predictions, one needs to change the matter to radiation density ratio 
to get the right position for the peaks. This can be done by either deducting from the radiation 
density, or adding more matter to the GA model. The first option 
is highly restricted from the CMB temperature data
\cite{fixsen}. The second option can be done either through adding baryons or cold dark matter,
or both. Since the density of baryons is constrained through helium abundance ratio (see
e.g.~\cite{helium}), the only remaining option is to add cold dark matter to the theory.
This is also limited by the ratio of even to odd peaks in the CMB power spectra, but is
the last resort! The best fit value for the cold dark matter density in the GA theory,
using CMB data only, is: $\Omega_{\rm DM} h^2 = 0.147 \pm 0.004$.

After fixing the position of the peaks, one needs to get the right amplitude for the spectra.
The relative amplitude of the high-$\ell$ to low-$\ell$ multi-poles is highly affected by the
early ISW effect that was explained before and is evident in the left panel of Fig.~\ref{fig:cell}
by comparing the ratio of the power of the two curves in $\ell \sim 250$, and $\ell \sim 2000$.
This relative mismatch in the amplitude can be fixed by choosing higher values of the spectral index,
$n_s$. The best fit value of this parameter in the GA theory is: $n_s = 1.042 \pm 0.008$.

\begin{figure*}[ht]
\includegraphics[scale = 0.9]{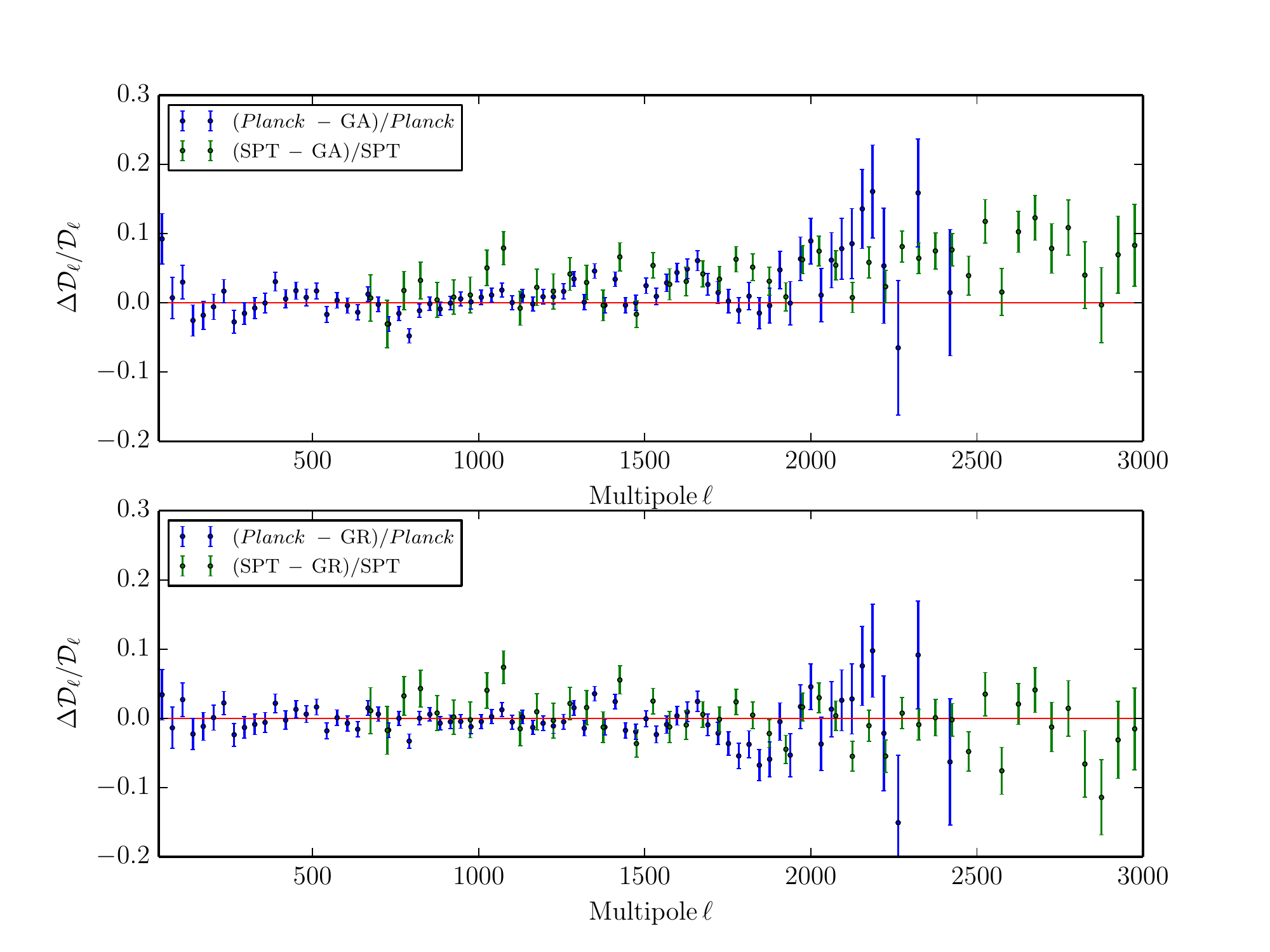}
\caption{\label{fig:errorbars} Comparing general relativity (bottom panel)
and gravitational aether (top panel)
predictions for the CMB power spectrum with $\Pl$ and SPT data sets. Here we plot ${\cal D}_\ell\equiv \ell(\ell+1)C_\ell1/2\pi$ residuals, along with
$\pm1\,\sigma$ error bars, from Refs.~\cite{Planck} and \cite{spt}.
While the two theories can both fit the lower-$\ell$ observations, GR fits 
the data points significantly better than GA for $\ell \gtrsim 1000$, at the
${>}\,4\,\sigma$ level.}
\end{figure*}

The best-fit predictions of the two theories are compared in the right panel of Fig.~\ref{fig:cell}.
We see that the best fit GA theory predicts less power at high $\ell$s compared to GR. The best-fit
predictions of the two theories are compared with $\Pl$ and SPT data in Fig.~\ref{fig:errorbars}.

\subsection{Cosmological constraints}
We now turn to deriving precision constraints on GGA from cosmological observations. We
assume that $\gn$ is equal to the Newtonian gravitational constant measured in Cavendish-type experiments (see e.g.\~Ref.~\cite{Willbook})
using sources with negligible pressure. Then {\tt CosmoMC} can be used for sampling $\gr$ using different combinations of the following 
cosmological data.
\begin{enumerate}
\item The first data release of the all-sky CMB temperature anisotropy power spectrum, measured by the \Pl~\cite{Planck1} satellite.\footnote{\url{http://pla.esac.esa.int/pla/aio/planckProducts.html}}
\item The 9-year (and final) data release of the {\it WMAP\/} satellite CMB temperature and polarization anisotropy power spectra, which we denote as \wmap-$9$~\cite{wmap} (with ``WP'' indicating the large angle polarization data only).
\item Three seasons of high resolution CMB temperature anisotropy measurements from the ACT experiment \cite{act}.
\item $790\,{\rm deg}^2$ of high resolution CMB temperature anisotropy measurements from the SPT experiment ~\cite{spt}.
\item Sloan Digital Sky Survey (SDSS)~\cite{sdss} and other estimates of the BAO length scale ~\cite{DR7,DR9,6DF}.
\item The first claimed detection of the amplitude of primordial gravitational waves, based on B-mode polarization anisotropy band-powers detected by the BICEP$2$ experiment at degree scales \cite{bicep}. 
\end{enumerate}

There are two special cases of particular interest, which are $\gr = \gn$
(standard General Relativity; $\zeta_4=0$) and $\gr = \frac{4}{3} \, \gn$ (Gravitational Aether
theory; $\zeta_4=\frac{1}{3}$). If the data are consistent with the $\gr / \gn = 4/3$ case, or favour this 
theory over GR, then that would be evidence that GA theory provides a better 
description of the cosmological data.

\begin{table*}
\footnotesize 
\nointerlineskip
\setbox\tablebox=\vbox{
\newdimen\digitwidth 
\setbox0=\hbox{\rm 0} 
\digitwidth=\wd0 
\catcode`*=\active 
\def*{\kern\digitwidth} 
\newdimen\signwidth 
\setbox0=\hbox{{\rm +}} 
\signwidth=\wd0 
\catcode`!=\active 
\def!{\kern\signwidth} 
\newdimen\pointwidth 
\setbox0=\hbox{\rm .} 
\pointwidth=\wd0 
\catcode`?=\active 
\def?{\kern\pointwidth} 
\halign{\hbox to 1.20in{#\leaderfil}\tabskip=2.0em&
\hfil#\hfil\tabskip=2.0em&
\hfil#\hfil\tabskip=2.0em&
\hfil#\hfil\tabskip=2.0em&
\hfil#\hfil\tabskip=0.0em\cr
\noalign{\doubleline}
Parameter& \wmap-$9$& WP$\,+\,$\Pl& WP$+$\Pl$+$HighL& WP$+$\Pl$+$BAO\cr
\noalign{\vskip 3pt\hrule\vskip 4pt} 
$ \Omega_{\rm b} h^2$&{$0.0226 \pm 0.0005$}&{$0.0227 \pm 0.0005$}& $0.0225 \pm 0.0004$&$0.0223 \pm 0.0003$\cr
$\Omega_{\rm DM} h^2$&$0.14 \pm 0.03$&$0.128 \pm 0.006$&$0.124 \pm 0.005$&$0.125 \pm 0.005$\cr
$100\, \theta$&$1.038 \pm 0.003$&$1.0421 \pm 0.0008$&$1.0418 \pm 0.0007$&$1.0415 \pm 0.0006$\cr
$\tau$&  $ 0.088 \pm 0.014$&$0.097 \pm 0.015$&$0.096 \pm 0.015$&$0.090 \pm 0.013$\cr
${\rm log}(10^{10} A_{\rm s})$& $3.10 \pm 0.04$&$3.11 \pm 0.03$&$3.10 \pm 0.03$&$3.10 \pm 0.03$\cr
$n_{\rm s}$& $0.979 \pm 0.019$&$0.987 \pm 0.017$&$0.975 \pm 0.014$&$0.970 \pm 0.009$\cr
$\gn / \gr $&$0.86 \pm 0.18$&$0.913 \pm 0.048$&$0.951 \pm 0.043$&$0.959 \pm 0.035$\cr
\noalign{\vskip 3pt\hrule\vskip 4pt}
}}
\endtable
\caption{Mean likelihood values together with the
 $68\%$ confidence intervals for the usual six cosmological parameters
 (see Ref.~\cite{Planck}), together 
 with the GGA parameter $\gn/\gr$. ``WP'' refers to \wmap-9 
 polarization, which has been used to constrain the optical depth, $\tau$.
 ``HighL'' refers to the higher multipole data sets, ACT and SPT.
 The PPN parameter, $\zeta_4$ can be obtained through
 $\zeta_4=G_{\rm R}/G_{\rm N}-1$.}
    \label{table:tbl1}
\end{table*} 

\begin{table*}
\footnotesize 
\nointerlineskip
\setbox\tablebox=\vbox{
\newdimen\digitwidth 
\setbox0=\hbox{\rm 0} 
\digitwidth=\wd0 
\catcode`*=\active 
\def*{\kern\digitwidth} 
\newdimen\signwidth 
\setbox0=\hbox{{\rm +}} 
\signwidth=\wd0 
\catcode`!=\active 
\def!{\kern\signwidth} 
\newdimen\pointwidth 
\setbox0=\hbox{\rm .} 
\pointwidth=\wd0 
\catcode`?=\active 
\def?{\kern\pointwidth} 
\halign{\hbox to 1.20in{#\leaderfil}\tabskip=2.0em&
\hfil#\hfil\tabskip=2.0em&
\hfil#\hfil\tabskip=2.0em&
\hfil#\hfil\tabskip=0.0em\cr
\noalign{\doubleline}
Parameter& WP$+$\Pl$+$BICEP2& WP$+$\Pl$+$HighL$+$BICEP2& WP$+$\Pl$+$BAO$+$BICEP\cr
\noalign{\vskip 3pt\hrule\vskip 4pt} 
$ \Omega_{\rm b} h^2$&$0.0229 \pm 0.0005$& $0.0228 \pm 0.0004$&$0.0223 \pm 0.0003$\cr
$\Omega_{\rm DM} h^2$&$0.132 \pm 0.006  $&$0.128 \pm 0.005   $&$0.127 \pm 0.005$\cr
$100\, \theta$&$1.0425 \pm 0.0008$&$1.0422 \pm 0.0007 $&$1.0416 \pm 0.0006$\cr
$\tau$&$ 0.101 \pm 0.015 $&$0.101 \pm 0.015   $&$0.090 \pm 0.013$\cr
${\rm log}(10^{10} A_{\rm s})$&$3.11 \pm 0.03    $&$3.11 \pm 0.03     $&$3.10 \pm 0.03$\cr
$n_{\rm s}$&$1.001 \pm 0.016  $&$0.991 \pm 0.015   $&$0.976 \pm 0.009$\cr
$r$&$0.18 \pm 0.04    $&$0.18 \pm 0.04     $&$0.16 \pm 0.03$\cr
$\gn / \gr$& $0.871 \pm 0.045  $&$0.905 \pm 0.040   $&$0.938 \pm 0.034$\cr
\noalign{\vskip 3pt\hrule\vskip 4pt} 
}}
\endtable
\caption{Mean likelihood values together with the
 $68\%$ confidence intervals for the usual six cosmological parameters, plus
 $r$ (the tensor-to-scalar ratio), together with the GGA parameter $\gn/\gr$.
 The data are as in Table~\ref{table:tbl1},
 but now including BICEP2 measurements of the B-mode CMB polarization.
 GR is still favoured over GA if we include HighL CMB or BAO measurements.
 However, even the conventional seven parameter
 GR model (that includes $r$), is disfavoured at 
 around the $3\,\sigma$ level when one considers BICEP2, as well as
 $\Pl$, and HighL data. The PPN parameter, $\zeta_4$ can be obtained through
 $\zeta_4=G_{\rm R}/G_{\rm N}-1$.}
    \label{table:tbl2}
\end{table*} 

\begin{figure*}[ht]
\includegraphics[scale = 1.5]{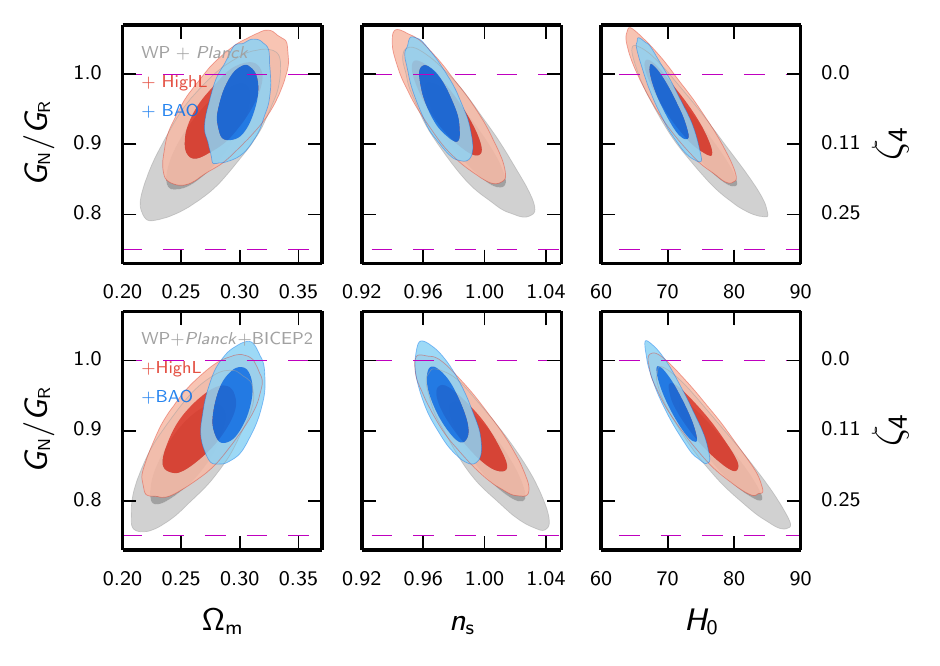}
\caption{\label{fig:degeneracy} Confidence intervals ($68 \%$ and $95 \%$)
for the GGA parameter and the cosmological parameters it is most degenerate
with.  The ratio $G_{\rm N}/G_{\rm R}$ is plotted on the left axes and
$\zeta_4$ on the right axes.  The horizontal dashed lines indicate
the GR (top line) and GA (bottom line) predictions.}
\end{figure*}

From a broader perspective, \textit{any} unequal values for $\gn$
and $\gr$ would be interesting, because this is a way of parameterizing general deviations from the matter-radiation equivalence principle.
The MCMC constraints on GGA, excluding the recent BICEP$2$ data release, are summarized in Table ~\ref{table:tbl1}. 
It is important to allow the usual cosmological parameters to vary while constraining $\gn /\gr$. 
This is because there could be (and indeed are) degeneracies in the new $7$-parameter (or $8$-parameter when the tensor-to-scalar ratio $r$ is
included) space. Some of these degeneracies between the GGA parameter, $\gn /\gr$, and the conventional
parameters of cosmology are shown in Fig.~\ref{fig:degeneracy}.

\begin{figure}[ht]
\includegraphics[scale = 0.45]{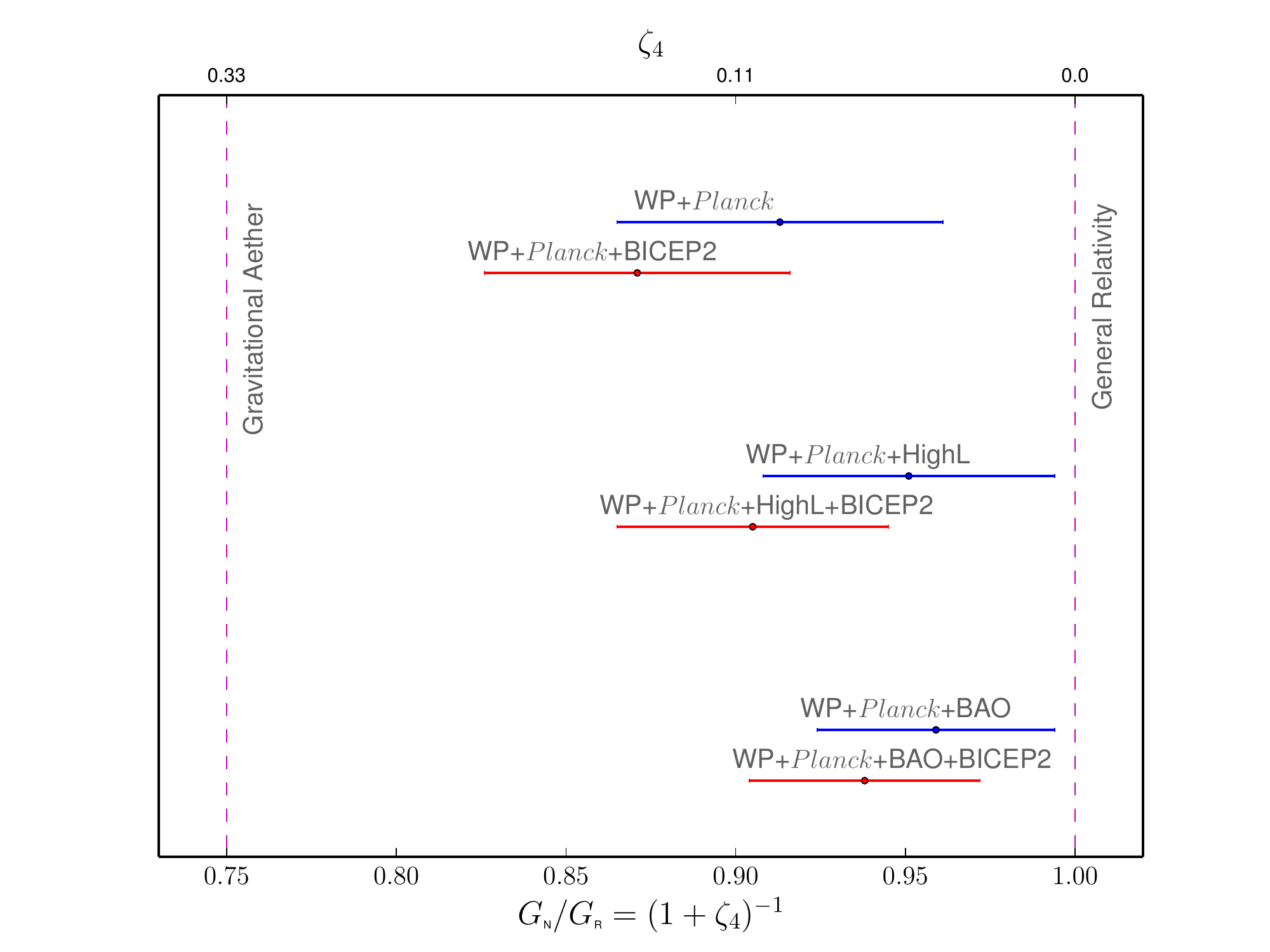}
\caption{\label{fig:planck_like} A pictorial comparison of marginalized $\gn / \gr = (1+\zeta_4)^{-1}$ measurements.  We have plotted the central values and
$\pm1\,\sigma$ error bars using different data sets. The GR and GA predictions are shown as
vertical dashed lines.}
\end{figure}

In fact, we find that if one omits the BICEP2 data, then $\gn /\gr =1$ provides a good fit and the cosmological parameters hardly 
shift from their best-fit GR values.  On the other hand, adding BICEP$2$ data shifts the results towards GA by about $1\,\sigma$.  This may be pointing to
some tension in data, or a mild inconsistency between GR and the existing
data sets.  Of course the most exciting possibility that any such tension
is due to missing physics rather than systematic effects.
Table ~\ref{table:tbl2} shows these constraints, while
Fig.~\ref{fig:planck_like} presents a pictorial comparison of constraints on
$\gn / \gr$ using different data sets.

\section{Discussion}

As we can see in Fig.~\ref{fig:planck_like}, although GR is generally
preferred over GA, different combinations of data sets appear to give
constraints for the GGA parameter (or anomalous pressure coupling), which are
discrepant by as much as 2$\,\sigma$.  Perhaps most intriguingly, the
combination of {\it Planck\/} temperature anisotropies
and polarization from {\it WMAP}-9 and BICEP2 (which represents the
state of the art for CMB anisotropy measurements above $0.1^\circ$), lies
about mid-way between the GA and GR predictions (with a preference for GA,
but only at the level of $\Delta\chi^2\,{\simeq}\,1$).
Nevertheless, the best fit for $G_{\rm N}/G_{\rm R}$ is inconsistent with both GA and GR at 2.7 and 2.9$\,\sigma$, respectively. The latter is a manifestation of the
well-known tension between the {\it Planck\/} upper limit on tensor modes, and
the reported detection by BICEP2 (at least for standard $\Lambda$CDM cosmology
with a power-law primordial power spectrum).

Let us now try to qualitatively understand what might be responsible for the
different trends that we observe when fitting different data sets, as we turn
up the GGA parameter.  The first step is to obtain the gross structure of the
CMB $C_\ell^{\rm TT}$ power spectrum peaks by fixing $\theta$, the ratio of the
sound horizon at last scattering, to the distance to the last-scattering
surface. For any value of $G_{\rm N}/G_{\rm R}$, this can be done by picking
appropriate values of $\Omega_m$ and $h$, which explains the degeneracy
directions in Fig.~\ref{fig:planck_like} for these parameters.

The next step is to recognize the effect of free steaming on the damping tail
of the CMB power spectrum. Similar to the effect of free streaming of additional
neutrinos, boosting the gravitational effect of neutrinos leads to additional
suppression of power at small scales, or high $\ell$, in the CMB power spectrum,as we can see in Fig.~\ref{fig:errorbars}. This can be partially compensated
for by increasing the spectral index of the scalar perturbations, leading to a
bluer primordial spectrum.  In fact, we see that combinations of data sets
that prefer larger $G_{\rm R}$ (in Tables~\ref{table:tbl1}--\ref{table:tbl2})
prefer a near scale-invariant power spectrum, $n_{\rm s}\simeq1$
(which is up from the value $n_{\rm s}\simeq 0.96$ in GR+$\Lambda$CDM).

Finally, a bluer scalar spectral index tends to suppress scalar power for
$\ell \lesssim 100$, which then relaxes the upper bound on tensors from the
{\it Planck\/} temperature power spectrum.   This allows a higher value of
$r$ than the limit ($r<0.11$ \cite{Planck}) found from the temperature
anisotropies in $\Lambda$CDM.

Of course, none of these degeneracies are perfect.  In particular, the
additional damping due to free-streaming is much steeper than a power law,
which is why even the best-fit GA model underpredicts CMB power for
$\ell \gtrsim 1000$ in Fig.~\ref{fig:errorbars}. This is also why adding higher
resolution CMB observations (from ACT and SPT), pushes the best fit away from
GA.  It is possible that adding a positive running for the spectral index might be able to partially cancel the effect, at least for the observable range of
multipoles.  However, a significant positive running would be hard to justify
in simple models of inflation, and may also exacerbate the observational
tensions with structure formation on small scales in $\Lambda$CDM.  

A more stringent constraint on GA (and thus anomalous pressure coupling) comes
from the degeneracy with the Hubble constant, which can also be seen in
Fig.~\ref{fig:degeneracy}.  Additional gravitational coupling to pressure,
of the sort required in GA, requires
$H_0 > 80\,{\rm km}\,{\rm s}^{-1}\,{\rm Mpc}^{-1}$, which is larger than even
the highest measurements in the current literature (see e.g., figure~16 in
Ref.~\cite{Planck}).
In particular, BAO geometric constraints place tight bounds of
$H_0 \simeq 68$--$72\,{\rm km}\,{\rm s}^{-1}\,{\rm Mpc}^{-1}$, which is why
the inclusion of these data substantially cuts off the smaller values of
$G_{\rm N}/G_{\rm R}$.  However, we should note that this inference is based on a simple
cosmological constant model for dark energy at low redshifts; more complex
descriptions of dark energy, as suggested by some recent BAO
\cite{2014arXiv1404.1801D} or Supernovae~Ia \cite{2013arXiv1310.3828R} studies,
could relax these $H_0$ constraints.   

There are certainly hints of possible systematics among the different data sets
that could explain some of these tensions.  For example, the power spectrum of
{\it WMAP}-9 appears to be about 2.5\% higher than {\it Planck\/}
\cite{Planck,2013MNRAS.436.1422K}, independent of scale.  Additionally, the
first 30 or so multipoles appear low (in both {\it WMAP\/} and {\it Planck\/}
data), which, coupled with calibration, can affect the best fit in the damping
tail.  A perhaps related issue is that the 
best-fit lensing amplitude in {\it Planck\/} and {\it Planck\/}+HighL spectra,
appears to be around 20\% higher than expected in the $\Lambda$CDM model
\cite{Planck, 2014arXiv1403.4599B}.  Since lensing moves power from small
$\ell$s to high $\ell$s, this could also have an indirect effect on the shape
of the high-$\ell$ power spectrum.

Finally, there are legitimate questions about whether BICEP2 analysis
\cite{bicep} has underestimated the effect of instrumental systematics or
Galactic foregrounds (e.g., \cite{2014arXiv1404.1899L}).  Decreasing the
primordial amplitude of B-modes would reduce the tension with {\it Planck},
and thus relax the need for anomalous pressure coupling (i.e.,
$G_{\rm N} < G_{\rm R}$).

On balance it seems premature to claim that $\zeta_4>0$ is required by the
current cosmological data.  The simple GA theory (with $\zeta_4=1/3$)
certainly appears disfavoured by the data.  However, as the quality of the
data continue to improve, it is worth bearing in mind that the GGA picture
provides a particular degree of freedom.  This should be considered in
future fits, particularly with the upcoming release of the {\it Planck\/}
polarization data.

\section{Conclusions, and Open Questions}\label{sec:sum}

In this paper, we have closely examined the question of anomalous pressure
coupling to gravity in cosmology.  This was done in the context of the
Generalized Gravitational Aether framework, which allows for an anomalous
sourcing of gravity by pressure ($\zeta_4$ in the PPN framework), while not
affecting other precision tests of gravity.  The idea would mean that the
gravitational constant during the radiation era, when $p=\frac{1}{3}\rho$,
is boosted to $G_{\rm R}=(1+\zeta_4)G_{\rm N}$, compared to the gravitational
constant for non-relativistic matter $G_{\rm N}$.  In particular, the case
with $\zeta_4=1/3$ or $G_{\rm R}=4G_{\rm N}/3$, can be used  to decouple vacuum
energy from gravity, and thus solve the (old) cosmological constant problem.

We have implemented cosmological linear perturbations for this theory into
the code {\tt CAMB}, and explored the models that best fit different
combinations of cosmological data.  The effects are qualitatively similar to
introducing additional neutrinos ($N_{\rm eff}$), or dark radiation.
Our constraints are summarized in Tables~\ref{table:tbl1}--\ref{table:tbl2}
and Figs.~\ref{fig:degeneracy}--\ref{fig:planck_like}.

There is clearly some mild tension between different data combinations,
but $\zeta_4=1/3$ is inconsistent with current observations at around the
2.6-$-5\,\sigma$ level, depending on the combination used.  CMB B-mode
observations (from BICEP2) push for larger $\zeta_4$, while high resolution
CMB or baryonic acoustic oscillations, go in the opposite direction.
The best fit is in the range $0.04 \lesssim \zeta_4 \lesssim 0.15$,
or $ 0.87 \lesssim G_{\rm N}/G_{\rm R} \lesssim 0.96$, with statistical errors
of a half to third of this range.  It may be interesting to notice that
even GR ($\zeta_4=0$) is disfavoured at $3\,\sigma$ when we combine 
lower resolution CMB observations.

To bring some statistical perspective, we should note that even if the
gravitational aether solution to the cosmological constant problem is ruled
out at $5\,\sigma$, the standard GR+$\Lambda$CDM paradigm, {\it with no
fine-tuning}, is ruled out at ${>}\,10^{60}\,\sigma$!  Therefore, while the
first attempt at solving the problem might not have been entirely
successful (compared to a model that takes the liberty of fine-tuning the
vacuum energy), we argue, that it may be a step in the right direction.
So, other than working to improve the quality and consistency of
observational data, what can we do to tackle this problem, that quantum
fluctuations appear not to gravitate?

From the theoretical standpoint, there are several clear avenues that we have
already alluded to:
\begin{enumerate}

\item As we discussed in the Introduction, gravitational aether is a classical
theory for an effective low energy description of gravity.  Therefore, like all
effective theories, it has an energy cut-off above which it will not be valid.
In fact, the length-scale $\lambda_{\rm c}$ (inverse energy scale) associated
with this cut-off should be $\lambda_{\rm c} \sim 0.1$ mm, since a smaller
$\lambda_{\rm c}$ would not fully solve the cosmological constant problem,
while larger $\lambda_{\rm c}$ could have been seen in torsion balance tests
of gravity (although it is not entirely clear what the signature would be).
It is worth noting that the number density of baryons at CMB last scattering
is $0.33\,{\rm mm}^{-3}$, implying that to calculate $T^\alpha_\alpha$
in Eq.~\ref{eqn1}, it might be necessary to use a microscopic description of
atoms interacting with aether, as opposed to the usual mean fluid density
picture \footnote{The density of dark matter particles is much more model
dependent, but is expected to be even less than
this baryon value for conventional WIMP models}.  If this is the case, then
each microscopic particle would carry an aether halo of size about
$\sim \lambda_{\rm c}$;  this would appear like a renormalization of particle
mass for all macroscopic gravitational effects, but otherwise (like for other
vacuum tests of gravity), the theory would be indistinguishable from GR.
Nevertheless, in lieu of a quantum theory of gravitational aether, it is not
clear how much progress can be made in this direction.

\item Another possibility is to modify the simple ansatz (\ref{tprime}) for
the energy-momentum tensor of the gravitational aether, e.g., by introducing
a density, $\rho'$.  This might be a reasonable approach if one is also
attempting to connect gravitational aether to dark energy (which does have
both density and pressure at late times).  However, Eq.~\ref{constraint} will
no longer be sufficient to predict the evolution of the aether, and thus we
would need another equation to fix the aether equation of state. 

\item In solving for the evolution of aether with respect to dark matter,
$\omega$, we have assumed that the two substances were originally comoving,
i.e., $\omega=0$ at early times.  However, depending on the process that
generates primordial scalar fluctuations in this picture, $\omega$ could have
also been sourced in the early Universe.  So, even though its amplitude decays
as $a^{-1}$ on super-horizon scales, depending on its amplitude and spectrum,
it can impact CMB observations.  This would be akin to introducing isocurvature
modes, but for aether perturbations.  Although, since $\omega$ decays
exponentially on sub-horizon scales, this could only affect the CMB at
$\ell \lesssim 100$. 

\item Finally, we have not included the effect of neutrino mass in our GGA
treatment.  Massive neutrinos will be qualitatively different from other
components, as they start as radiation, which does not couple to aether, but
then gradually start sourcing aether as they become non-relativistic.  However,
this happens relatively late in cosmic history, long after CMB last-scattering,
and when neutrinos make up only a small fraction of cosmic density.
Therefore, although this would be a useful direction to pursue, we do not
expect a significant change from the analyses presented here.    

\end{enumerate} 

In contrast to unfalsifiable approaches for solving the cosmological constant
problem, such as landscape/multiverse ideas with anthropic arguments, the
gravitational aether concept has the very distinct advantage of being
predictive and hence it can be falsified.  Here, we have demonstrated this
explicitly, since the basic picture does not appear to fit the current
cosmological data.
However, like elsewhere in physics, the logical next step would be to learn
from this process and propose better physical models (rather than relying
on metaphysics).  We believe that the GGA approach yields a useful
parameterization of a particular degree of freedom in models of modified
gravity, and that this idea is worth pursuing further.

\section*{Acknowledgement}
This research was enabled in part by support provided by WestGrid
(www.westgrid.ca) and Compute Canada Calcul Canada
(www.computecanada.ca). We would like to thank Siavash Aslanbeigi for his useful comments.
NA is supported by the Natural Science and Engineering 
Research Council of Canada, the University of Waterloo and Perimeter Institute for 
Theoretical Physics. Research at the Perimeter Institute is supported by the Government 
of Canada through Industry Canada and by the Province of Ontario through the Ministry 
of Research \& Innovation. AN would like to thank Jeremy Heyl and Ariel Zhitnitsky for 
useful comments and discussion. 

\newpage
\bibliography{GA_cosmology_4}

\end{document}